\documentclass[aps,pra,twocolumn,showpacs,epsfig]{revtex4}
\usepackage{dcolumn}
\usepackage{bm}
\usepackage{graphicx}
\usepackage{amsmath}
\usepackage{latexsym}
\usepackage{bbold}
\usepackage{color}
\usepackage{amsfonts}
\usepackage{amssymb}
\usepackage{array}
\usepackage{times}
\usepackage{epsfig}
\usepackage{epstopdf}
\usepackage{setspace}
\newcommand{\ket}[1]{\left\vert#1\right\rangle}

\newcommand{\nbar}{\overline{n}}

\date{\today}
\begin{document}
\title{The role of environmental correlations in the non-Markovian dynamics of a spin system}
\author{Salvatore Lorenzo$^{1,2,3}$, Francesco Plastina$^{1,2}$, and Mauro Paternostro$^3$}
\affiliation{$^1${Dipartimento di  Fisica, Universit\`a della Calabria, 87036 Arcavacata di Rende (CS), Italy}\\
$^2$ INFN - Gruppo collegato di Cosenza \\
$^3$Centre for Theoretical Atomic, Molecular and Optical Physics, School of Mathematics and Physics, Queen's University, Belfast BT7 1NN, United Kingdom}
\begin{abstract}
We put forward a framework to study the dynamics of a chain of
interacting  quantum particles affected by individual or
collective multi-mode environment, focussing on the role played by
the environmental quantum correlations over the evolution of the
chain. The presence of entanglement in the state of the
environmental system magnifies the non-Markovian nature of the
chain's dynamics, giving rise to structures in figures of merit
such as entanglement and purity that are not observed under a
separable multi-mode environment. Our analysis can be relevant to
problems tackling the open-system dynamics of biological complexes
of strong current interest.
\end{abstract}


\maketitle

The investigation on the interaction between a spin-like two-level
system (TLS) and an environment embodied by an ensemble of
harmonic oscillators, the so-called spin-boson
model~\cite{LeggettReview}, has gained renewed momentum due to the
current strong interest in the quantum-mechanical description of
excitation-transfer processes in biological organisms or the
dynamics of protein-solvent systems in photosynthetic complexes
~\cite{Huelga,Reina,lloyd,scholak,Huelga2}. The interest in such a
celebrated model is twofold, in this context. On one side, it is
employed to understand the active role of an environment in
assisting transport across such structures~\cite{lloyd}. On the
other hand, the model is extremely useful to characterize the
extent to which non-classical behaviors survive in systems
affected by environmental worlds that are strictly adherent to the
paradigm of being {\it wet and warm}. In this respect, we aim at
gathering a description of the TLS-environment evolution that is
as complete as possible and able to account for subtle yet
potentially key features of such dynamics emerging from the
finiteness the environment itself, its capability to retain
information on the system it is coupled to (and kick it back at
later times) and its inherent structure. For,
instance, Rebentrost {\it et al.} have shown in Ref.~\cite{nmqj}
that environmental memory-keeping effects are crucial for
preserving the coherent behavior of excitonic energy transfer
processes in biological complexes.

Stepping aside from interesting yet still controversial
bio-physical scenarios, the general model addressed above plays an
important role in various contexts of applied quantum physics,
from cavity quantum electrodynamics to its solid-state analogues
involving defects in diamond coupled to semiconducting
microcavities~\cite{diamond} or superconducting devices and planar strip-line
resonators~\cite{cqed}. In the latter case, very interesting regimes are quickly
becoming available where the coupling strength of the coherent
part of an evolution can be made comparable to the one responsible
for the open-system dynamics due to the presence of the
environment \cite{bourassa}.

All such considerations and physical configurations demand the
development of a framework apt to encompass and harness features
of non-Markovianity in various coupling regimes. Formally, one is
required to retain, to some extent, the degrees of freedom of the
environment so as to allow for its possibly-non-trivial structure
to influence the dynamics of the system of interest. This can be
computationally very demanding and while sophisticated
techniques have been put forward to determine the
quantum statistical properties of interacting quantum many-body
systems, they are either hardly suitable for non-linear
configurations or face some difficulties in tracking the temporal dynamics of the system itself.

In this paper we contribute to the investigation on the
interaction between a system of interacting TLS's coupled to a
bosonic environment by studying the effect that environmental
entanglement has on the dynamics of a chain of interacting TLS's.
We build up a mathematical apparatus that, inspired by and
significantly extending the approach in Ref.~\cite{massimo},
solves exactly the {\it local} interaction between a single TLS
and its multi-mode environment, so as to build an eigenbasis that
is then used in order to tackle the inter-particle coupling by
keeping track of the environmental properties. This helps us in
incorporating the full quantum features of the environmental
state, therefore enabling a quantitative study of the role of
entanglement in the evolution of the chain under investigation and
its relation to the non-Markovian nature of the dynamics. We
illustrate our techniques by addressing the cases of a dimer and a
trimer, which are the smallest non-trivial configurations allowing
us to study the most crucial attributes of such a complex problem.
Our investigation opens up the possibility to study full-scale
networks of coupled TLS's in an open-system fashion, covering
the full range of coupling strengths with the environments, from
the weak to the strong one. Although inevitably leaving a few
questions open, we believe such an endeavor to be helpful in the
contexts described above.

The remainder of this paper is organized as follows: in
Sec.~\ref{modelmethod} we describe the method put forward in order
to analyze the dynamics of an $N$-site chain interacting with
individual multi-mode bosonic environments without resorting to
the weak-coupling regime and thus allowing for a more complete
overview of the effects of non-Markovianity on the evolution of
the chain. Sec.~\ref{diff} is devoted to the explicit study on a
dimer of interacting particles affected by entangled environments.
The key features in the dynamics of such configuration are
compared to what is achieved under the assumption of independent
environments so as to show that entanglement appear to enhance the
non-Markovian aspects of the time evolution. Sec.~\ref{comm}
addresses the case of a common environment affecting the sites of
our dimer, showing that the manifestations mentioned above are
insensitive to the non-local nature of the environment, but rather
depend on the properties of its state such as the
entanglement-sharing structure, as studied in Sec.~\ref{3} for the
case of a trimer. Finally, in Sec.~\ref{conc} we draw our
conclusions and briefly discuss the questions opened by our
investigation.

\section{The model and the analytical method to its approach}
\label{modelmethod} The model under consideration is described by
an Hamiltonian of the form
\begin{equation}
\label{htot}
\mathbf{\hat H_{tot}=\hat H_S+\hat H_{SS}+\hat H_B+\hat S_{SB}}.
\end{equation}
Here $\mathbf{\hat H_S}=-\hbar\sum_{j=1}^N
\epsilon_j\hat\sigma_{j}^z$ gives the free dynamics of the
elements of the chain, $\epsilon_j$ being the energy split between
the levels of each TLS, while
$\{\hat\sigma_j^x,\hat\sigma_j^y,\hat\sigma_j^z\}$ are the Pauli
matrices for particle $j{=}1,..,N$. Analogously, $\mathbf{\hat
H_B}{=}\sum_{j=1}^N\sum^{n_j}_{k=1}{\hbar \omega_{j,k} \hat
a_{j,k}^{\dag }\hat a_{j,k}}$ is the free energy of the
environmental harmonic oscillators. The pedex $k$ labels the $n_j$
independent modes interacting with the TLS at site $j$. Each of
such modes has frequency $\omega_{j,k}$ and is described by the
bosonic annihilation [creation] operator $\hat a_{j,k}$ [$\hat
a_{j,k}^\dag$]. On the other hand, $\mathbf{\hat H_{SS}}$ models
the inter-particle coupling within the chain, while
$\mathbf{\hat H_{SB}}$ describes the TLS-environment interaction.
We take these two terms as
\begin{equation}
\label{hshss}
\begin{aligned}
 \mathbf{\hat H_{SB}}=&\sum_{j=1}^N \hat\sigma_{j}^z\otimes\sum^{n_j}_{k=1} g_{j,k}(\hat a_{j,k}^\dag + \hat a_{j,k}),\\
 \mathbf{\hat H_{SS}}=&-2\hbar\sum_{n=j}^{N-1}J_{j,j+1}(\hat \sigma_{j}^x\hat\sigma_{j+1}^x+\hat\sigma_{j}^y\hat\sigma_{j+1}^y)
\end{aligned}
\end{equation}
with $J_{j,j+1}$ being the nearest-neighbor coupling strength
between the particles occupying sites $j$ and $j+1$ on the chain
and $g_{j,k}$ the analogous parameter associated with the
interaction of the $j^{\text{th}}$ TLS and its $k^{\text{th}}$
environmental mode.

The first of Eq.~\eqref{hshss} generates a displacement of the
environmental oscillators, conditioned on the state of the
corresponding TLS. Our approach to the description of the dynamics
is to solve the problems embodied by each the subsystem made out
of a single TLS and the ensemble of local modes it interacts with.
We thus consider $\mathbf{\hat H}_\mathbf{0}^j=\mathbf{\hat
H}_\mathbf{S}^j+\mathbf{\hat H}_\mathbf{B}^j+\mathbf{\hat
H}_\mathbf{SB}^j $ and look for eigenstates of the form $| a
\rangle_j \otimes | \phi_i \rangle_j$ with $a=\pm$ and
$\{\ket{\pm}_j\}$ denoting the eigenbasis of $\hat\sigma^z_j$. States
$| \phi_{i} \rangle_j$ are the eigenstates of the total
Hamiltonian of an ensemble of oscillators (per assigned state of
the corresponding TLS), that is
\begin{equation}
\sum^{n_j}_{k=1}[\pm  g_{j,k} (a_{j,k}^{\dag}{+}a_{j,k}) + \hbar\omega \hat a_{j,k}^{\dag} \hat a_{j,k}]\!|\phi_{\pm} \rangle_j{=}(E^j\pm\epsilon_j) | \phi_{\pm} \rangle_j,
\end{equation}
where $E^j$ is the total energy of the $j^{\text{th}}$ subsystem. A detailed calculation shows that such energy depends on the number of excitations $n_{j,k}$ in the oscillators coupled to the TLS at the $j^{\text{th}}$ site according to the expression
\begin{equation}
E^j_n =\pm\epsilon_j+ \sum^{n_j}_{k=1}\hbar\omega_{j,k}(n_{j,k} -{\alpha_{j,k}^2}),
\end{equation}
where $\alpha_{j,k}=g_{j,k}/\hbar\omega_{j,k}$. The corresponding
eigenstates can be written as $|\psi_{\pm}^{{\bf
n}_j}\rangle_j{=}|\pm\rangle_j{\otimes}|\alpha_{\pm}^{{\bf
n}_j}\rangle_j{=}|\pm\rangle_j{\otimes^{n_j}_{k=1}}|\alpha_{\pm}^{n_k}\rangle_j$,
where ${\bf n}_j=\{n_1,n_2,..,n_{n_j}\}$ is the vector specified
by the number of excitations populating the $n_j$ oscillators
interacting with the $j^{\text{th}}$ TLS and we have introduced
the excitation-added coherent states $
|\alpha_{\pm}^{n_k}\rangle{=}[{(a^{\dag}_j\pm\alpha_{j,k})^{n_k}}/{\sqrt{n_k
!}}]|\pm\alpha_{j,k}\rangle$. Although it is straightforward to
check that $\langle \alpha_+^m |\alpha_+^n \rangle{=}\delta_{mn}$
($\langle \alpha_-^m|\alpha_-^n \rangle{=}\delta_{mn}$) and the
set of states $\{| +,\alpha_{+}^n \rangle\}$ ($\{| -,\alpha_{-}
^n\rangle\}$) form an orthonormal basis, states $| \alpha_{+}^n
\rangle$ and $| \alpha_{-}^n \rangle$ are not mutually
orthogonal. A treatment with an arbitrary number of modes per site, performed
along the lines of the discussion above, is certainly possible
although computationally demanding. Therefore, from now on, we
will focus our attention to the case of a single mode per site and
drop the mode-label.\\
In the oscillator basis discussed above, the Hamiltonian (\ref{htot}) takes the form
\begin{equation}
\label{htot2} \mathbf{\hat H_{tot}}=\sum_{j=1}^N\mathbf{\hat
H}_\mathbf{0}^j+\mathbf{\hat H_{SS}},
\end{equation}
each $\mathbf{\hat H}_\mathbf{0}^j$ being now diagonal. The
remaining term $\mathbf{\hat H_{SS}}$, which is defined in the
second line of Eq.~\eqref{hshss}, only contains the degrees of
freedom of the TLS chain. In the following, we consider a
homogeneous chain with $J_{j,j+1}=J,~\forall{j}=1,..,N$. By
introducing the raising and lowering operators
$\hat\sigma^\pm_{j}=(\hat\sigma^x_j\pm i\hat\sigma^y_j)/2$, 
it is convenient to rewrite it as
\begin{equation}
\label{hss2}
 \mathbf{\hat H_{SS}}=-\hbar J\sum_{n=1}^{N-1}(\hat\sigma_{j}^+\hat\sigma_{j+1}^-+\hat\sigma_{j}^-\hat\sigma_{j+1}^+).
\end{equation}
In the interaction picture defined with respect to the unperturbed
Hamiltonian $\mathbf{\hat H_0}{=}\sum_j\mathbf{\hat
H}_\mathbf{0}^j$, we get
\begin{equation}
\mathbf{\hat H_I}(t){=}-\hbar
J\!\!\left(\prod_{j=1}^{N}\hat\Theta^\pm_j
\right)\!\sum_{n=1}^{N-1}(\sigma_{n}^+\sigma_{n+1}^-+h.c.)\!
\left(\prod_{k=1}^{N}\hat\Theta^\pm_{k}\right)^{\dag}
\end{equation}
with $ \hat \Theta^\pm_j=\sum_{n=0}^\infty{e^{i E_{\pm}^n
t}|\pm,\alpha_{\pm}^{n} \rangle\langle\pm,\alpha_{\pm}^{n}|}$.\\
After elaborating on this expression, we obtain
\begin{equation}
\label{hi} \mathbf{\hat H_I}(t) =-\hbar J\sum_{n=1}^{N-1}(\hat
\sigma_{n}^+\hat \sigma_{n+1}^- \hat\Theta^{+}_{n}\hat
\Theta^{-}_{n+1}\hat\Theta^{-\dag}_{n}\hat\Theta^{+\dag}_{n+1}+h.c.).
\end{equation}
By expanding the $\hat \Theta^{\pm}_{n}$'s in a coherent-state
basis, we finally arrive at the effective expression for the
intra-chain Hamiltonian
\begin{equation}
\label{finale} \mathbf{\hat H_I}(t) =-\hbar J\sum_{n=1}^{N-1}[\hat
\Gamma_{n}^+\hat  \Gamma_{n+1}^-+ \hat \Gamma_{n}^-\hat
\Gamma_{n+1}^+],
\end{equation}
where $\hat \Gamma^\pm_j=\hat
\sigma^\pm_j\otimes\hat D_j[\pm\alpha(t))]$ with $\hat{D}(\xi)$
the displacement operator of amplitude
$\xi\in\mathbb{C}$~\cite{barnettradmore} and
$\alpha_j(t)=2\alpha_j(1-e^{i\omega_j t})$.
While the analysis above explicitly consider individual
environments affecting the elements of the chain, one can also
study the complementary situation where all the TLS's collectively
interact with a single (and common) ensemble of oscillators. In
this case, $[\mathbf{\hat H_{tot}},\sum_{j}\hat\sigma_z^j]{=}0$,
which in turn implies that $\mathbf{H_{SS}}$ can be easily
diagonalized. We tackle an explicit example of this situation in
Sec.~\ref{comm}.

In order to obtain information on the full dynamics of the TLS
subsystem, we write the density matrix $\rho(t)$ of the
chain-oscillators system in the excitation-added displaced basis.
For all our simulations,  we assume the factorized initial state
\begin{equation}
\rho(0)=\rho_\text{chain}(0) \otimes \rho_\text{ho}(0)
\end{equation}
with $\rho_\text{chain}(t)$ [$\rho_\text{ho}(t)$] describing the
state of the chain (oscillator environment) at time $t$. In what
follows we consider two significant instances of
$\rho_\text{ho}(0)$: the tensor product
of individual thermal states and the case of entangled oscillators
prepared in $N$-mode squeezed states~\cite{barnettradmore}. As it
will be discussed later on, such three paradigmatic cases allow us
to well identify the effect of the entanglement among the
environments on the dynamics of the TLS's.

Let us start addressing the case of a thermal state
\begin{equation}
\rho_{\text{th}}=\underset{j=1}{\overset{N }{\bigotimes }}\sum_{l=0}^\infty\dfrac{\nbar^l_j}{(1+\nbar_j)^{l+1}}|l\rangle_j\langle l|,
\end{equation}
where $\nbar_j{=}(e^{\beta\hbar \omega_j}{-}1)^{-1}$ is the average number
of quanta in the oscillator at temperature $T$ (we have taken the
{\it inverse temperature} $\beta=1/k_b T$ with $k_b$ the
Boltzmann constant) and $\ket{l}$ is the state with $l$
excitations. In the excitation-added displaced basis of the
oscillators, we can recast $\rho_{\text{th}}$ into the form
\begin{equation}
\rho_{\text{th}}{=}\underset{j=1}{\overset{N }{\bigotimes }}\sum_{n,m,l=0}^\infty\dfrac{\nbar^l_j}{(1+\nbar_j)^{l+1}}
|\alpha_{\pm}^{n}\rangle_j\langle\alpha_{\pm}^{n}|l\rangle_j\langle l|\alpha_{\pm}^{m}\rangle_j\langle\alpha_{\pm}^{m}|,
\end{equation}
where the $+$ or $-$ sign for the $j^\text{th}$ oscillator has to be chosen according to the state of the corresponding element in $\rho_\text{chain}$.
The scalar products $\langle\alpha_{\pm}^{n}|l\rangle$ needed for the evaluation of the expression above are given in terms of Charlier polynomials $C_n(m; \alpha^2)$ as~\cite{AS,comment}
\begin{equation}
\begin{aligned}
\langle\alpha_{-}^{n}|m\rangle&=\dfrac{\alpha^{n+m}e^{-\alpha^2/2}}{\sqrt{n!}\sqrt{m!}}C_n(m; \alpha^2),\\
\langle\alpha_{+}^{n}|m\rangle&=\dfrac{\alpha^{n+m}e^{-\alpha^2/2}}{\sqrt{n!}\sqrt{m!}}C_m(n; \alpha^2),
\end{aligned}
\end{equation}
where we have assumed $\alpha\in\mathbb{R}$. In the remainder of the manuscript we take $\nbar_j=\nbar,\,\,\forall{j}$. We will also consider
an $N$-mode a pure squeezed-vacuum state for the harmonic
oscillators
$\ket{\text{sq}_N}{=}(1{-}\lambda^2)^{-1/2}\sum_l\lambda^l\otimes^N_{j=1}\ket{l}_j$
with $\lambda=\tanh r$ and $r$ the squeezing parameter. This
state can be expressed in the chosen oscillator basis following
the very same lines highlighted above and using again the Charlier
polynomials. It is worth emphasizing that these two environmental states are
locally indistinguishable as a squeezed vacuum is locally
equivalent to a thermal state.
\section{Chain dynamics under independent environments}
\label{diff}

We are now in a position to start addressing the effective
dynamics  of the TLS chain under the influences of the various
instances of environment introduced in the previous Section. In
order to strip down our discussion from unnecessary complications,
we study here the smallest configuration of the model at hand,
i.e. a two-TLS chain (a {\it dimer}) affected by two individual
harmonic oscillators, one per site. The two TLS will be considered
as identical. While embodying an interesting enough
situation~\cite{Reina}, this case allows for an agile description
of the physics involved.

In the dimer basis, the interaction hamiltonian in
Eq.~(\ref{finale})  takes the form
 \begin{equation}
\mathbf{\hat H_I}(t)=
\left(\begin{array}{cccc}
 0 & 0 & 0 & 0 \\
 0 & 0 & \hat D_1(\alpha_1)\hat D^\dag_2(\alpha_2) & 0 \\
 0 &\hat D^\dag_1(\alpha_1)\hat D_2(\alpha_2) & 0 & 0 \\
 0 & 0 & 0 & 0  \\
\end{array}
\right)
\end{equation}
In order to simplify the notation, we take identically coupled
TLS-oscillator subsystems, so that $\alpha_j=\alpha~(j{=}1,2)$
(this assumption, as well as the one of identical TLS's, does not
limit the generality of our results and can be relaxed with only
mild complications). The corresponding time-evolution operator
$\hat{\cal U}(t)$ is then calculated as a time-ordered Dyson
series $\hat{\cal U}(t)=\sum_{m}\hat{U}_n(t)$ with
\begin{equation}
\hat U_m(t){=}{\frac{(-i)^m}{m!}}\!\!\int_{0}^t{\!\!dt_1}\!\!
\int_{0}^t{\!\!dt_2}\cdot\cdot\!\!\int_{0}^t{\!\!dt_m\hat{\cal
T}\mathbf{\hat H_I}(t_1)\mathbf{\hat
H_I}(t_2){\cdot\cdot}\mathbf{\hat H_I}(t_m)},
\end{equation}
where $\hat{\cal T}$ is the time-ordering operator. The
anti-diagonal form of $\mathbf{\hat H_I}(t)$ helps considerably in
working out a manageable expression or the time-evolution
operator. In fact, the even (odd) terms in the
expansion of $\hat{\cal U}(t)$ involve diagonal (anti-diagonal)
operators, in the dimer basis. In details, we get
\begin{equation}
\begin{aligned}
&\underset{k=1}{\overset{2m}{\prod }}\mathbf{\hat H_I}(t_{k}){=}|+-\rangle_{12}\langle+-|{\otimes}\!\underset{k=1}{\overset{m}{\prod }} \hat{\cal D}_{+-}(\alpha(t_{2k-1}))\hat{\cal D}_{+-}^\dag(\alpha(t_{2k}))\\
&{+}|-+\rangle_{12}\langle-+|{\otimes}\!\underset{k=1}{\overset{m}{\prod}}\hat{\cal D}_{-+}^\dag(\alpha(t_{2k-1}))\hat{\cal D}_{-+}(\alpha(t_{2k})),\\
&\underset{k=1}{\overset{2m+1}{\prod }}\mathbf{\hat H_I}(t_{k}){=}|+-\rangle_{12}\langle-+|{\otimes}\!\underset{k=0}{\overset{m}{\prod }}\hat{\cal D}_{+-}(\alpha(t_{2k}))\hat{\cal D}_{-+}^\dag(\alpha(t_{2k+1}))\\
&{+}|-+\rangle_{12}\langle+-|{\otimes}\!\underset{k=0}{\overset{m}{\prod }} \hat{\cal D}_{-+}^\dag(\alpha(t_{2k}))\hat{\cal D}_{+-}(\alpha(t_{2k+1}))
\end{aligned}
\end{equation}
where $\hat{\cal D}_{+-}(\alpha(t_k)){=}\hat D_1(+\alpha(t_k))\hat
D_2^\dag(-\alpha(t_k))$  [with an analogous definition for
$\hat{\cal D}_{-+}(\alpha(t_k))$].
\begin{figure}[t!]
\center{{\bf (a)}\hskip4cm{\bf (b)}}
\center{\includegraphics[width=4.3cm]{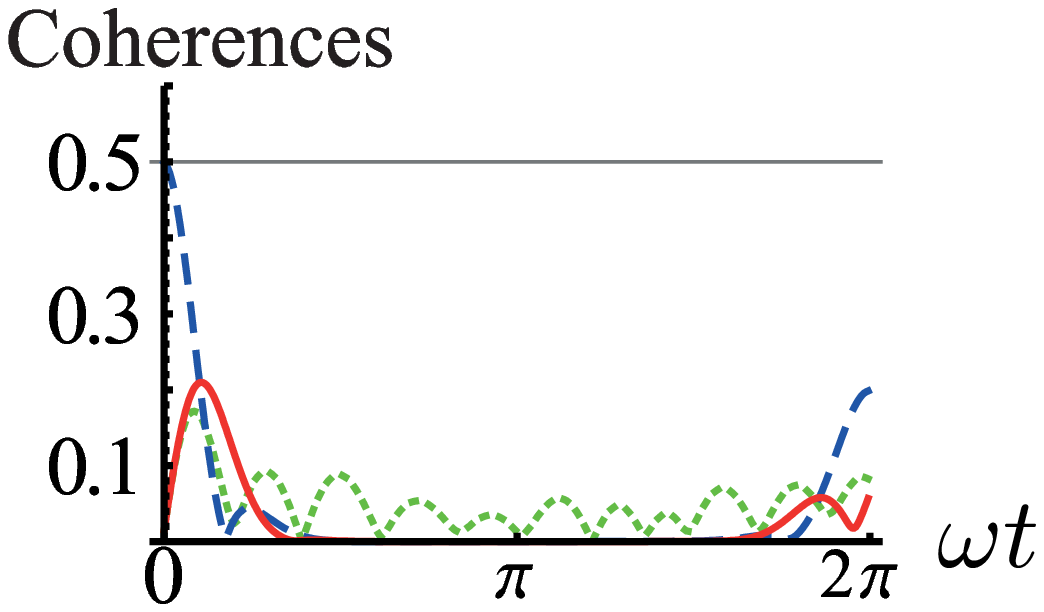}\includegraphics[width=4.3cm]{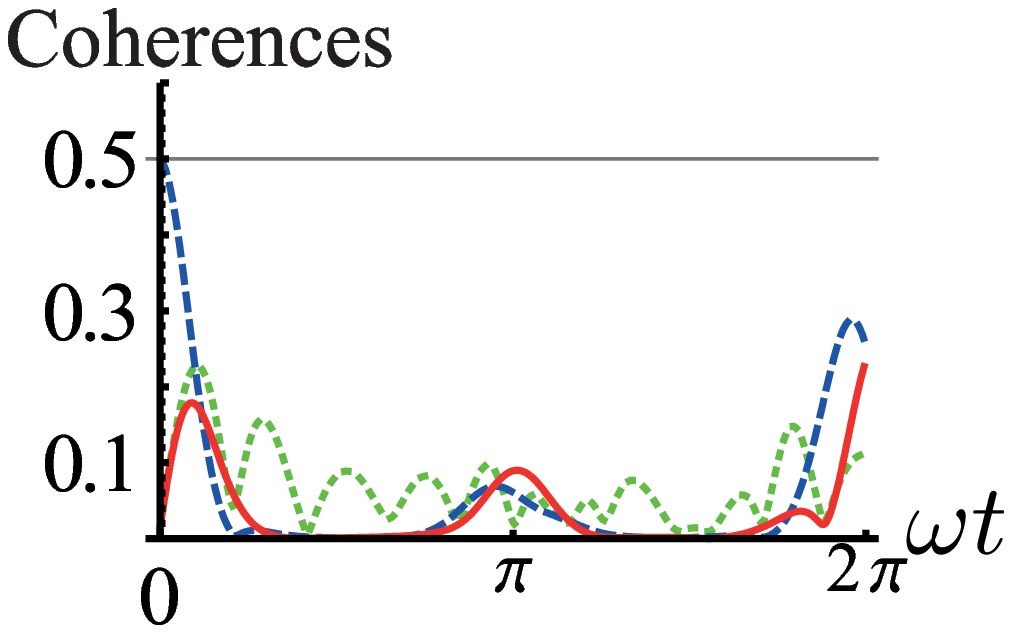}}
\caption{(color online): Coherences of the dimer's density matrix
for a chain affected by an environment prepared in a thermal and a
squeezed-vacuum state [panel {\bf (a)} and {\bf (b)},
respectively]. We have taken identical parameters for the two
TLS-oscillator subsystems with $\hbar J=0.04$eV, $\hbar
\omega=0.02$eV, $g=0.02$eV at room temperature  ($T=293$ K). In
this case $\alpha=1$. As explained in the text, the squeezing
factor for the results in panel {\bf (b)} has been taken so that
the reduced single-oscillator state has the same effective
temperature as in the thermal state case [i.e. panel {\bf (a)}].
Quantitatively, $r= 0.816$. A coherence revival at $t=\pi/\omega$
is present for a squeezed-vacuum environment: the presence of
quantum correlations between the environmental modes fundamentally
affects the dynamics of the TLS's.}
   \label{coh-term-sque}
\end{figure}

By initializing the first (second) TLS in the coherent
superposition $\ket{\text{in}}_1{=}a|-\rangle_1+b|+\rangle_1$ with
$|a|^2+|b|^2{=}1$ ($\ket{\text{in}}_2=\ket{-}_2$) and considering
that $[\mathbf{H_{tot}},\sum^2_{j=1}\hat\sigma^z_j]=0$, the state
$|+,+\rangle_{12}$ will stays unpopulated, while the probability
to populate $|-,-\rangle_{12}$ remains  constant. In
Fig.~\ref{coh-term-sque} we plot the coherences of the density
matrix of the two TLS's affected by boson environments prepared in
either uncorrelated thermal states or a squeezed-vacuum one. In
order to perform a faithful comparison between such two cases, we
have taken the squeezing parameter according to $\tanh
r=e^{-\frac{\beta\hbar\omega}{2}}$ with the value of $\beta$
provided by the thermal-state environment. This ensures that the
reduced single-oscillator state obtained from the squeezed-vacuum
environment is a thermal state having the same temperature as the
one considered in the genuinely thermal instance. In this way, we
can nicely isolate the effect of the entanglement shared by the
environmental oscillators.

\begin{figure}[b!]
\centering
\includegraphics[width=6cm]{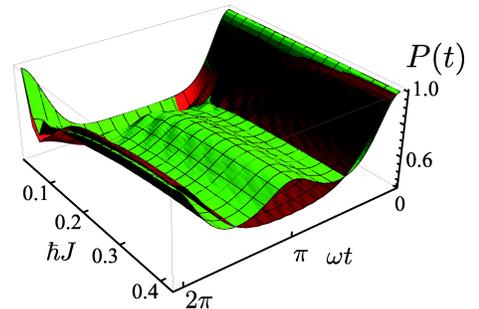}
\caption{(Color online): Purity of dimer's density matrix affected
by individual environments prepared in a thermal state (red
surface) and a squeezed-vacuum one (green surface, staying above
the previous one). $P(t)$ is plotted against the chain inter-site
coupling energy $\hbar J$ for $h\omega=0.04$eV, $g=0.02$eV and
$T=293$K, which corresponds to $\alpha=1/2$. We have taken a
squeezing factor $r=0.489$ (see text for further details on this
choice).}\label{J-function}
\end{figure}

Clearly, the revival of coherences at $t={\pi}/{\omega}$ is only
due to the presence of correlations in the environment and an
indication of a significant deviation of the state properties in
such two different environmental preparations.
Fig.~\ref{J-function} shows the purity
$P(t)=\text{Tr}(\rho_{12}^2)$ of the chain's density matrix
$\rho_{12}$ in both the thermal and squeezed case. We plot such a
figure of merit against the intra-site coupling strength $J$ and
the re-scaled interaction time $\omega{t}$. Quite reasonably, at
values of $J$ that are comparable with the TLS-oscillator
coupling, the differences between the thermal and squeezed-vacuum
cases becomes very small. We can understand this by considering
that, for decreasing values of $J$, the non-local interaction
between the sites of the chain becomes less relevant than the
local TLS-oscillator coupling. This implies that each TLS is
strongly tied to its associated mode so that the local features of
the oscillator environment become more preponderant than the
non-local ones and the differences with a configuration of
independent environmental states fades away. On the contrary, the
inter-site tunnelling arising from a large value of $J$ makes the
inter-mode entanglement relevant. Such differences manifest
themselves in a rather distinct behavior of the properties of the
TLS state.

As a special but very interesting case, we consider the chain to
be initialized in the maximally entangled state ${({1}/{\sqrt
2})(|--\rangle+|++\rangle)_{12}}$ and wonder about the behavior
that entanglement has under the interaction model we are studying.
Such an initial state is interesting on its own as, remarkably, it
is relatively straightforward to find closed analytical
expressions for the elements of the reduced TLS density matrix
$\rho_{12}$.
\begin{figure}[t!]
{\bf (a)}\hskip4cm{\bf (b)}
\includegraphics[width=4.25cm]{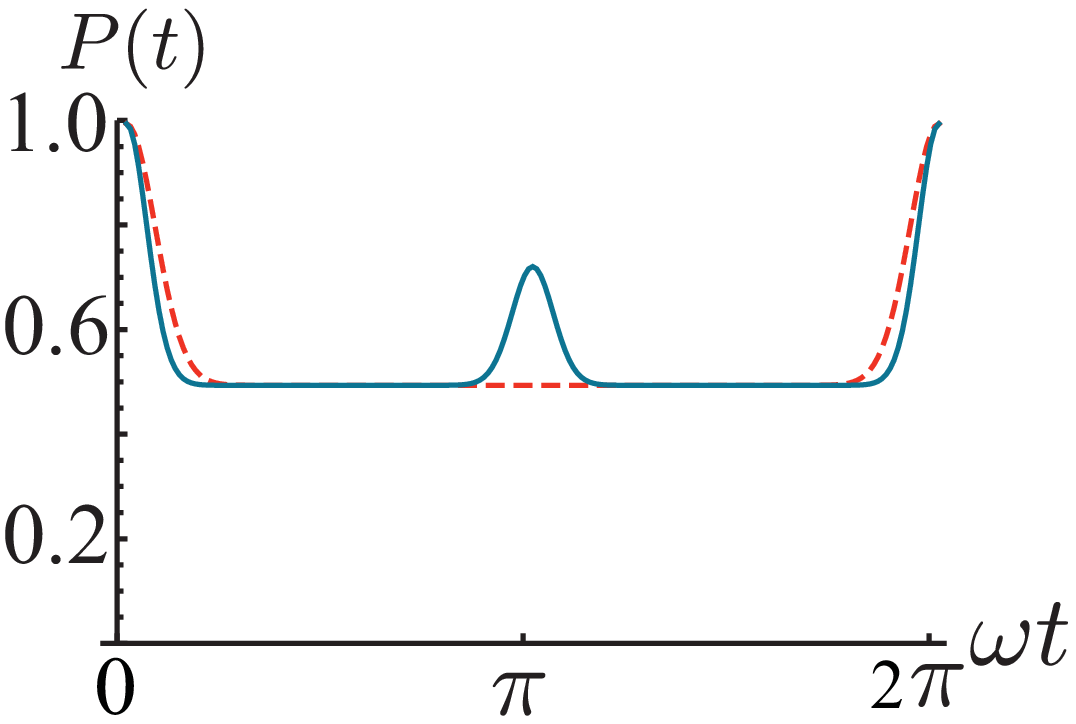}
\includegraphics[width=4.25cm]{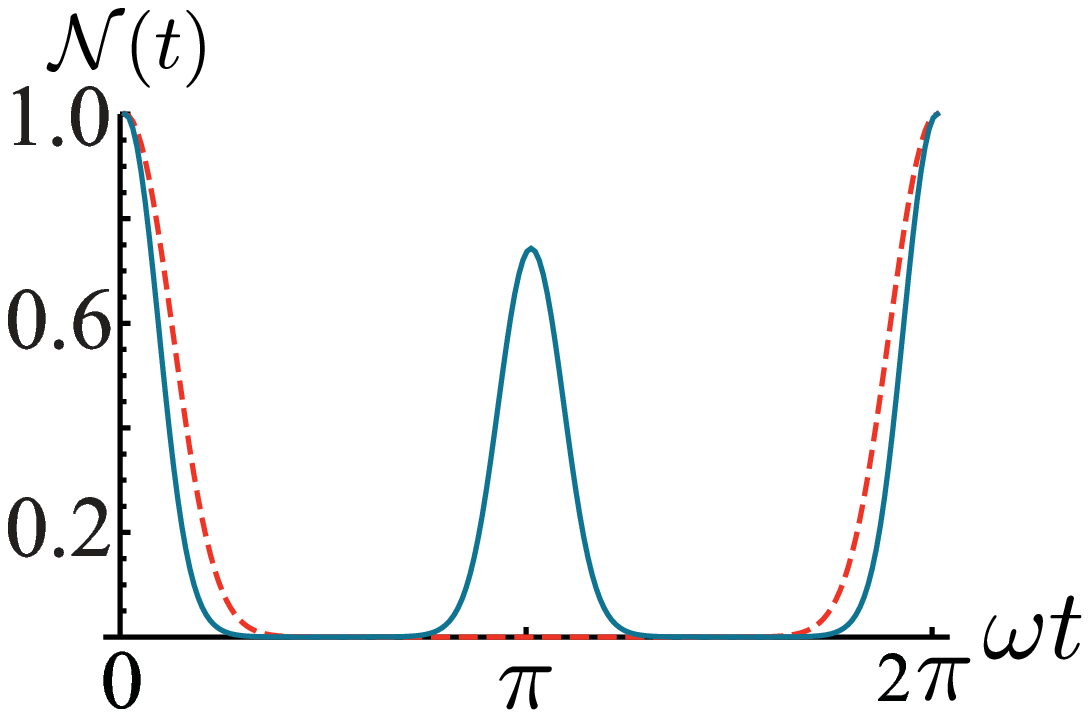}
\caption{(Color online): Purity [panel {\bf (a)}] and logarithmic
negativity [panel {\bf (b)}] of the state of a two-site chain
initially prepared in the maximally entangled state $({1}/{\sqrt
2})(|--\rangle+|++\rangle)_{12}$ and affected by either thermal
states of the oscillator environment (red dashed curve) or an
entangled squeezed-vacuum state (solid dark one) for $(\hbar
J,\hbar \omega,g)=(0.1,0.01,0.005)$eV and $r=1.158$,
corresponding to $\alpha=1/2$. The revival at $t=\pi/\omega$
associated with the squeezed-vacuum case is very
pronounced.}\label{bell}
\end{figure}
While the populations of such matrix remain constant, the
interesting element turns out to be the coherence
$\langle++|\rho_{12}(t)|--\rangle$, for which we find
\begin{equation}
\langle++|\rho_{12}(t)|--\rangle{=}\dfrac{1}{2}e^{4 i \epsilon t+
8\alpha
^2\frac{(\lambda_{\text{th}}^2+1)}{\lambda_{\text{th}}^2-1}[1-\cos(\omega
t)]t}
\end{equation}
for an oscillator environment prepared in the tensor product of
two individual thermal states, while
\begin{equation}
\label{sqcase} \langle++|\rho_{12}(t)|--\rangle{=}
\dfrac{1}{2}e^{4i\epsilon
t+8\alpha^2\frac{1{+}\lambda_{\text{sq}}^2{+}2
\lambda_{\text{sq}}\cos(\omega
t)}{\lambda_{\text{sq}}^2-1}[1{-}\cos(\omega t)]t}
\end{equation}
is valid for a squeezed-vacuum environment with
${\lambda_{\text{th}}=\lambda_{\text{sq}}^2=\nbar/(\nbar+1)}$.

The additional oscillatory term proportional to
$\lambda_{\text{sq}}\cos\omega{t}$ appearing in Eq.~(\ref{sqcase})
is responsible for the differences in the trend followed by the
coherences corresponding to such two cases. The state purity is
given by
\begin{equation}
P(t)=\dfrac{1}{2}+2|\langle++|\rho_S(t)|--\rangle|^2,
\end{equation}
which is reported in Fig.~\ref{bell} {\bf (a)} for the two
different types of environmental states considered in this study,
highlighting in a rather striking way how strongly the
entanglement between the environmental oscillators affects the
TLS's properties. The purity revival at $t=\pi/\omega$ is also
reflected in the behavior of the entanglement shared  by the
TLS's. We use logarithmic negativity~\cite{logneg} as our
entanglement measure. This is defined as ${\cal
N}(t)=\log_2||\rho^{PT}_{12}||$ with $\rho^{PT}_{12}$ the
partially transposed density matrix of the two TLS's and $||\hat
A||=\text{Tr}\sqrt{\hat A^\dag\hat{A}}$ the trace norm of
any operator $\hat{A}$. At $t=\pi/\omega$, ${\cal N}(t)$ goes from
zero (the system experiences a sudden death of entanglement)
to a rather large value.

\begin{figure}[t!]
      \includegraphics[width=6.3cm]{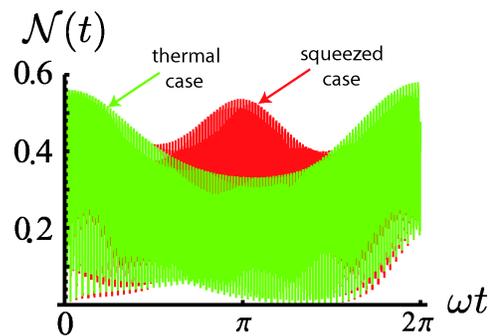}
\caption{(Color online): Logarithmic negativity between the two
sites of a chain affected by either thermal states of the
oscillator environment (green curve) or an entangled
squeezed-vacuum state (red one) for $r=0.816$ and $\alpha=1/4$. At
$t=\pi/\omega$, ${\cal N}(t)$ has a revival, for the
squeezed-vacuum case, which is absent from the thermal
state-affected system.}
  \label{logneg}
\end{figure}

The trend illustrated so far has features of independence from the
specific instance of initial state considered for the dimer. We
have studied the evolution of ${\cal N}(t)$ when the system is
initialized in a separable state, finding qualitatively analogous
behaviors of such figure of merit. For instance, Fig.~\ref{logneg}
reports the results corresponding to a chain prepared as $(1/\sqrt
2)(\ket{-}+\ket{+})_1\otimes\ket{-}_2$: entanglement builds up in
the system (both the curves in Fig.~\ref{logneg} are such
that ${\cal N}(0)=0$) and never disappears (we have considered a
relatively small coupling between environments and chain).
However, while the case corresponding to a thermal-state
environment has a periodicity of $2\pi/\omega$, when a squeezing
is present, a resurgence of entanglement occurs at half the
thermal-state period.

Such effects can be linked to an enhanced non-Markovian nature of
the TLS dynamics arising when considering entangled environmental
oscillators. The analysis of this connection is the main aim of
the following section.

\subsection{Characterizing the non-Markovian nature of the dynamics}

The quantification of the degree of non-Markovianity of adynamical
evolution has been the focus of a significant recent research
activity~\cite{meas1,meas2,meas3,meas4}. The main goal of such
investigation has been the identification of appropriate tools for
the characterization of the many facets of non-Markovianity and
the proposal of operational ways to quantify it. More recently,
such tools have found application in the characterization of the
memory-keeping evolution of systems of physical
relevance~\cite{app1,app2}. In this Subsection we move along such
lines and employ one of the above-mentioned measures to determine
the features of the dynamics under study here and the differences
between the two environmental configurations addressed so far.

Quantitatively, we take inspiration from the measure of
non-Markovianity proposed by Rivas {\it et al.} in
Ref.~\cite{meas2}, which is based on the following approach:
consider a bipartite entangled state such that only one component
of the bipartition is affected by the dynamics we would like to
characterize. Then, by computing the amount of entanglement
between the two parties at different instants of times within a
selected interval $[ t_0, t_\mathrm{max} ]$ we can detect
non-markovianity by checking for a non-monotonic behavior of the
quantum correlations. That is, for
$\rho_{SA}(0)=|\Phi\rangle\langle\Phi|$ with $|\Phi\rangle$, some
maximally entangled pure bipartite state, we define $\Delta
E=E[\rho_{SA}(t_0)]-E[\rho_{SA}(t_\mathrm{max})]$ (where $E$
denotes a legitimate entanglement monotone). We thus take
\begin{equation}
\label{huelga}
\mathcal{I}=\int_{0}^{t_\mathrm{max}}\left|\frac{dE[\rho_{SA}(t)]}{dt}\right|dt-\Delta E
\end{equation}
If the evolution of the system is Markovian, the derivative of
$E[\rho_{SA}(t)]$ is negative and $\mathcal{I}^{(E)}=0$.

While the measure in~\cite{meas2} has been explicitly formulated
for two qubits, so as to characterize the nature of a single-qubit
channel, one can certainly think about extending the argument so
as to address the case of a larger system. In particular, here we
would like to be able to qualitatively understand the reasons
behind the occurrence of the resurgence peak when an entangled
environment is at hand.

Without embarking into the challenge of a rigorous formulation of
the measure for non-Markovianity, we believe it is reasonable to
consider the following situation: we take a genuinely tripartite
entangled state of the two sites of the dimer at hand and an
additional, ancillary TLS, which is exposed to a perfectly unitary
dynamics (i.e. no environmental boson is interacting with the
ancilla). More specifically, we consider the W state
$(1/\sqrt3)(\ket{+--}+\ket{-+-}+\ket{--+})_{a12}$, which also
ensures the presence of entanglement in any two-TLS reduction.
While, as stated above, the ancilla $a$ undergoes a unitary
evolution, particles $1$ and $2$ are exposed to the effective
quantum channel described so far. 
\begin{figure}[t]
   \includegraphics[width=6cm]{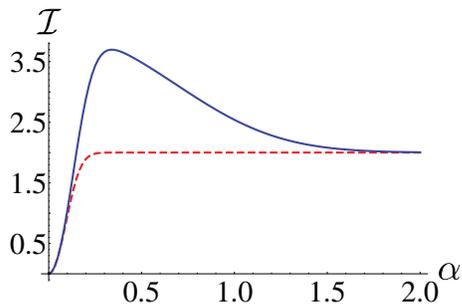}
\caption{(Color online): Witness of non-Markovianity for the thermal
(red dashed line) and squeezed blue solid line) case for $\hbar \omega=0.01$ and
$r=1.15$ against the coupling strength $\alpha$.}
  \label{huelgafig}
\end{figure}
We then apply
Eq.~(\ref{huelga}), where $E$ is taken to be the logarithmic
negativity between $a$ and the dimer, to the cases where the
environment is prepared in a squeezed-vacuum and a two-mode
thermal state. The results are shown in Fig.~\ref{huelgafig}
against the effective coupling rate $\alpha = g/\hbar \omega$ for
$t_\text{max}=2\pi/\omega$. While in the case of a thermal
environment, ${\cal I}$ grows monotonically and saturates for
$\alpha\gtrsim{0.4}$, for a squeezed-vacuum state it peaks at
about this value of $\alpha$, signaling an increased non-Markovian
behavior associated with such an environmental state. This
magnified non-Markovianity is at the origin of the recurrence
peaks exhibited in the analyses of the previous Sections. For
lager coupling strengths $\alpha$, each TLS-environment
interaction becomes so strong with respect to the inter-TLS one
that the system fragments into a series of {\it local}
sub-systems, each being basically oblivious to any non-local
nature of the environmental state.

\section{Chain dynamics under a common environment}
\label{comm}

We now move to the study of a configuration where the TLS chain is
affected by a common environment. The expression for the
system-environment coupling given in the first line of
Eq.~\eqref{hshss} is thus modified into
\begin{equation}
\mathbf{\hat H_{SB}}=g\sum^{n_o}_{k=1}(\hat{a}_k+\hat{a}^\dag_k)\otimes\sum^N_{j=1}\hat\sigma^z_j
\end{equation}
where we have assumed an oscillator environment consisting of
$n_o$ modes, each identically coupled to all the TLS of and
$N$-element chain. As remarked above, the symmetries in the
present case are such that one can easily gather the exact
analytical form of the evolved density matrix of the chain. For
illustrative purposes, and to make a clear comparison to the case
treated in previous Sections, we address a chain of two identical
element, this time affected by a common environment composed by
two boson modes. In Ref.~\cite{Reina}, an analogous problem has
been approached by restricting the analysis to the
single-excitation subspace and using a numerical method for the
analysis of non-Markovian dynamics. Here, we shall pursue an
analytic approach that is made possible by the analysis conducted
in Sec.~\ref{modelmethod}
\begin{figure}[b!]
 \includegraphics[width=7cm]{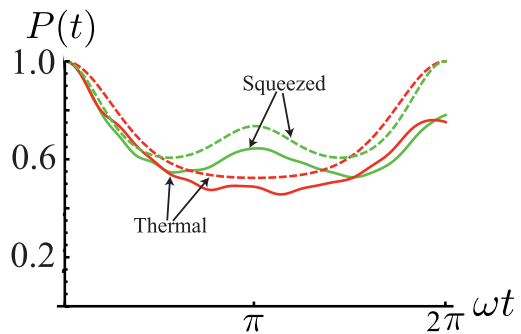}
\caption{(Color online) Purity against dimensionless interaction time $\omega{t}$
for thermal (red) and squeezed (green) environment states,  with
$\hbar J=0.1,\hbar \omega=0.03,g=0.01,\alpha=1/3$ . In the two
cases, squeezing parameter and (effective) temperature are
$r=0.62$ and $T=293$ K. The dashed curves refer to the case of 
common environments. } 
\label{common-diff}
\end{figure}
\begin{figure}[b!]
 \center{\bf (a)}\hskip4cm{\bf (b)}\\
   \includegraphics[width=4cm]{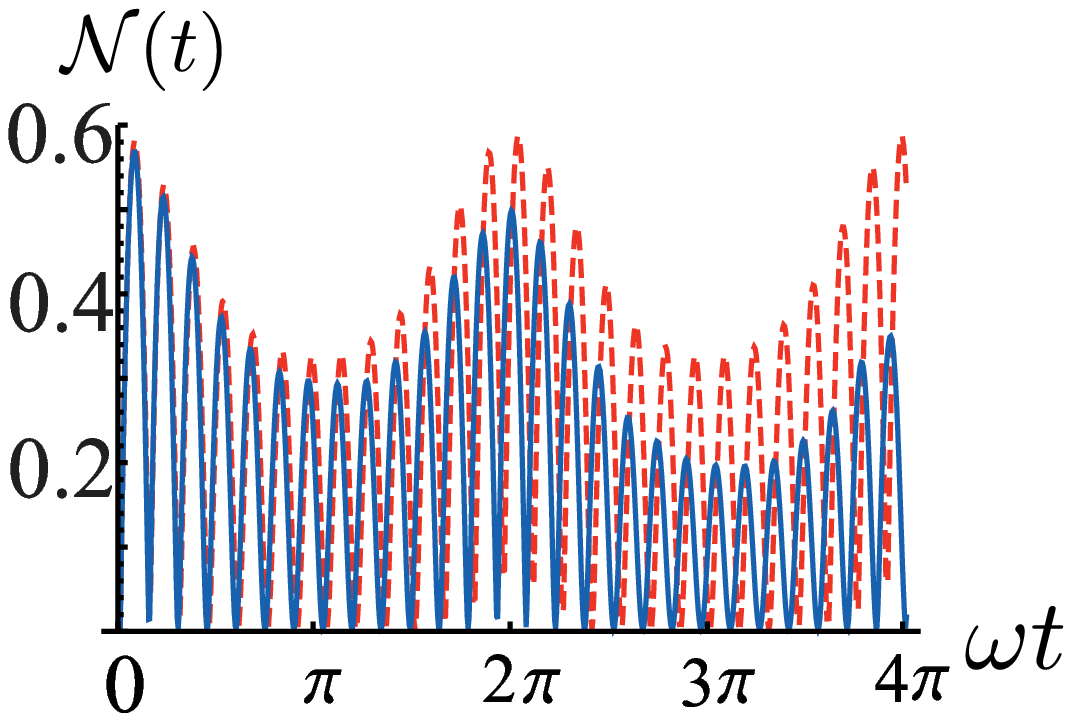}~~ \includegraphics[width=4cm]{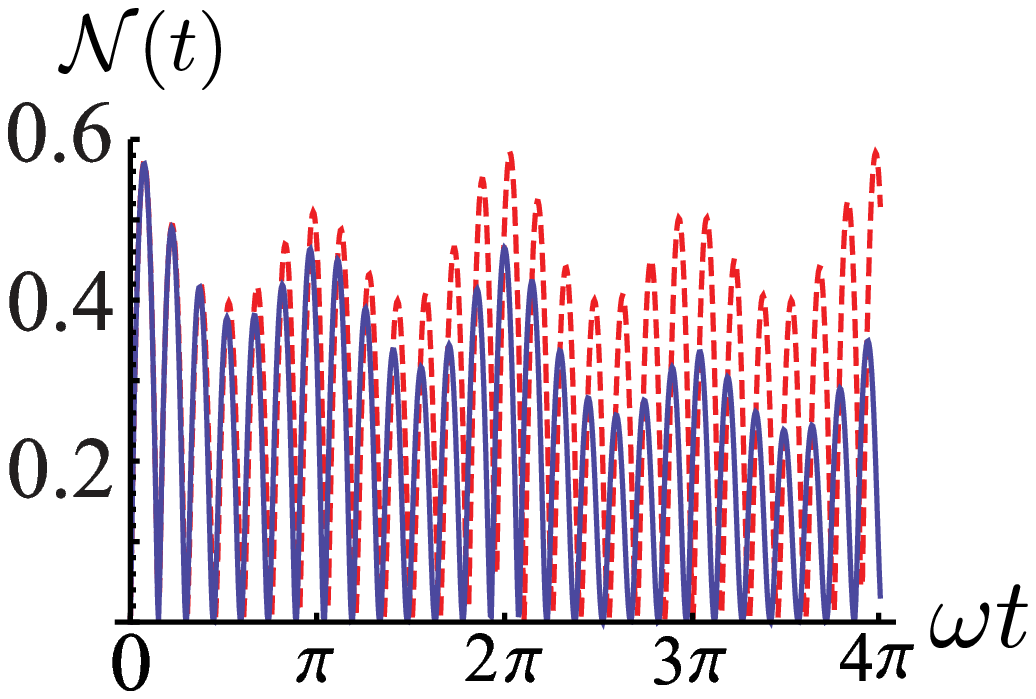}
\caption{(Color online): Logarithmic Negativity for $\hbar J/g=0.2$ and
$\alpha=1/3$, in the common- (dashed curve) and independent-
(solid line) environment case. Panel {\bf (a)} addresses the case of a thermal environment, while panel {\bf (b)} is for a squeezed one.}
\label{fig-varie}
\end{figure}
We start by considering an arbitrary pure state
$|\eta\rangle{=}{(a|-\rangle+b|+\rangle)}_1$ at site $1$ of the
chain, while the second element is prepared in $|-\rangle_2$. The
time evolution operator in the excitation-added displaced basis of
the environmental modes is given by
\begin{equation}
\begin{aligned}
&\hat{\cal U}(t)=\sum_{n,m}e^{i\omega n t}\left(f(t)^\pm |\pm\pm\rangle \langle \pm\pm|\otimes|\alpha^{\pm\pm}_{n,m}\rangle\langle\alpha^{\pm\pm}_{n,m}|\right.\\
&\left.+[\theta(t) |\mp\pm\rangle\langle\mp\pm|+h(t) |\mp\pm\rangle\langle\pm\mp|]\otimes|n,m\rangle\langle n,m|\right)
\end{aligned}
\end{equation}
with $\varphi^\pm(t)=e^{\pm i (2\epsilon+\omega\alpha^2)t}$, $\vartheta(t)=\cos(J t)$ and $\kappa(t)=i\sin(J t)$. Notice that, if $\alpha=0$, then $|\alpha^\pm_n\rangle=|n\rangle$.
With this expression at hand, the reduced density matrix of the chain is
\begin{widetext}
\begin{equation}
\label{grande}
\rho_{12}(t){=}
\begin{pmatrix}
 0 & 0 & 0 & 0 \\
 0 & b^2 \vartheta^2(t)&b^2 \vartheta(t)\kappa^*(t) & a b^*\varphi^+(t) \vartheta(t) \gamma(t)  \\
 0 & b^2 \vartheta(t)\kappa(t)&  b^2\kappa^2(t)    & a b^*\varphi^+(t) \kappa(t) \gamma(t)\\
 0 & a^* b \varphi^-(t)\vartheta(t)\gamma^*(t)  & a^* b \varphi^-(t)\kappa^*(t) \gamma^*(t)& a^2
\end{pmatrix}
\end{equation}
where we have introduced the {\it decoherence factor}
\begin{equation}
\gamma(t)=\exp[{{-2\alpha ^2\nbar(e^{i\omega t }-1)(e^{-i\omega t+\beta\omega }-1)}}]
\end{equation}
for the thermal state case and
\begin{equation}
\begin{aligned}
\gamma(t)=&\exp[{-2\alpha ^2\dfrac{(1-e^{-i \omega t })(1-e^{-i\omega t } \lambda_{\text{sq}} ^2)+[1-2 \cos(\omega t)+ \cos(2 \omega t)] \lambda_{\text{sq}}}{\lambda_{\text{sq}} ^2-1}}]
\end{aligned}
\end{equation}
for squeezed-vacuum one.
\end{widetext}

Very similar features between the individual-environment and the common-environment cases 
are found by studying the behavior of purity. Fig.~\ref{common-diff} shows that, regardless of the 
environmental configuration (whether made out of a common system or two individual oscillators), the squeezed-vacuum 
preparation is responsible for the revival of purity discussed above, a feature that is absent from the thermal case. 

It is interesting to notice that the central block of the density
matrix in Eq.~(\ref{grande}) does not contain any decoherence factor, thus being fully
decoherence-free: entanglement and purity will be preserved in
time, in such a subspace, which is associated with a single excitation. 
Clearly, this marks a net difference with the case of two independent environments, where no such
special symmetry exists. Such a distinction is evident from the inspection of Fig.~\ref{fig-varie}, where we plot the 
logarithmic negativity in the common and individual-environment cases, for both the thermal and squeezed-vaccum preparation: 
the long-time behavior in the individual-environment arrangement clearly manifests a decrease of ${\cal N}(t)$ that is 
absent from the common-environment configuration. 

\section{Three-site chain: the role of the sharing-structure of environmental entanglement}
\label{3}

After having analyzed in the previous Sections the basic
building-block for any complex structure of interacting TLS's, we
move to the  study of an open three-site chain (a trimer) affected
by individual environments. While, qualitatively, the result
gathered in Sec.~\ref{diff} will be fully confirmed, such a
scenario offers us an opportunity to investigate whether the
sharing structure of the environmental entanglement plays any role
in the open-system dynamics undergone by the TLS-chain.

\begin{figure}[ht!]
   \includegraphics[width=8cm]{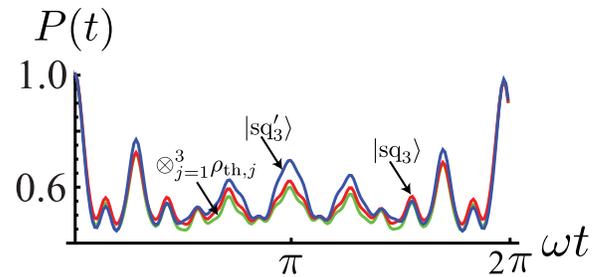}
\caption{(Color online): Purity of three sites chain for
the cases of thermal (green), generalized squeezed (red) and
double split (blue) environment states. The nature of genuine
tripartite entanglement of the generalized squeezed state is
almost invisible, while some differences in the time evolution can
be pointed out for the split squeezed state.}
\label{fig-varie1}
\end{figure}

\begin{figure*}[t!]
 \center{\bf (a)}\hskip5.5cm{\bf (b)}\hskip5.5cm{\bf (c)}\\
  \includegraphics[width=18cm]{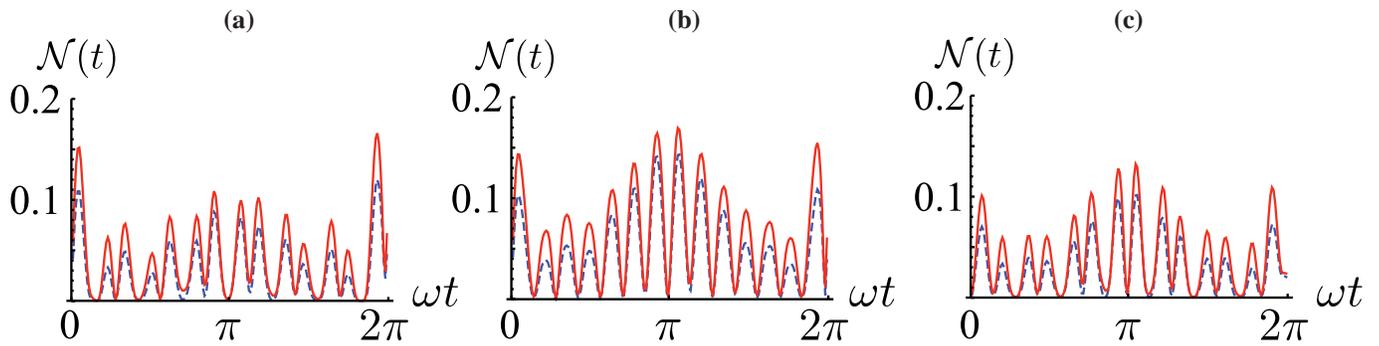}
\caption{Bipartite logarithmic negativity ${\cal N}(t)$ against $\omega{t}$ for the reduced state of two TLSs' obtained from a three-site chain by tracing out one of the particles. We have taken $\hbar{J}=0.2$eV, $\alpha=1/4$ with $r=0.488$ and $T=293$K. The solid (dashed) lines are for the squeezed (thermal) case.
Panel {\bf (a)}, {\bf (b)} and {\bf (c)} show the results achieved when particle $1$, $2$ and $3$ are traced out, respectively. 
}
\label{log-neg-3}
\end{figure*}

To start with, we have considered the three-mode squeezed state
$\ket{\text{sq}_3}{=}(1{-}\lambda^2)^{-1/2}\sum^3_{l=0}\lambda^l\ket{l,l,l}_{123}$.
This state is such that, upon tracing out one of the modes, the
remaining two are fully separable thermal states, thus witnessing
the GHZ-like nature of such state. We have found negligible
differences in the dynamics of the trimer arising from the use of
such environmental state or the tensor product
$\otimes^3_{j=1}\rho_{\text{th},j}$: as shown in
Fig.~\ref{fig-varie1}, for instance, the purity function
$P(t)$ of the three-TLS state corresponding to such cases is very
similar [as seen by inspecting the two bottom curves in
Fig.~\ref{fig-varie1}]. The insensitivity to squeezing is
confirmed by looking at the bipartite entanglement between any
pair of chain elements achieved after tracing out one
TLS-oscillator subsystem. Fig.~\ref{log-neg-3} 
illustrates the behavior of the logarithmic negativity for such
bipartite states; the results are valid, qualitatively, regardless
of the pair being taken and show that there is no revival peak at
$=\pi/\omega$ in the case of a squeezed environment. Moreover, a similar 
behavior is observed when considering the {\it tripartite negativity}, which is a 
legitimate entanglement monotone for the tripartite case~\cite{trineg}. 

The situation however changes if we modify the
entanglement-sharing structure among the environmental
oscillators. To show this fact, we have taken a three-mode state
obtained by superimposing at a beam-splitter one mode of a
squeezed state $\ket{\text{sq}_2}_{12}$ to an ancilla prepared in
the vacuum state $\ket{0}_3$. More formally, we consider the state
\begin{equation}
\ket{\text{sq}'_3}=(\openone_1\otimes\hat{B}_{23})(\theta)\ket{\text{sq}_2}_{12}\otimes\ket{0}_3
\end{equation}
with $\hat B(\theta)=\exp[({\theta/2})(\hat a^\dag_2\hat a_3-\hat
a^\dag_3\hat a_2)]$ being the beam splitter operator of
transmittivity $\cos^2(\theta/2)$~\cite{barnettradmore}. For
$\theta=0$ ($\theta=\pi$), a two-mode squeezed-vacuum state of
modes $1$ and $2$ ($1$ and $3$) is retrieved, while for
intermediate values of such parameter, an entangled three-mode
state is achieved having non-separable two-mode reduced states. In
this second case, the environmental entanglement visibly affects
the global properties of the system in a way consistent with what
we have seen earlier in Sec.~\ref{diff}, see Fig. \ref{fig-varie1}, thus reinforcing the idea that the structure of quantum
correlations shared by the environmental oscillators plays a key
role in the detailed non-Markovian dynamics of the TLS subsystem.

\section{Conclusions}
\label{conc}

We have studied the full non-Markovian dynamics of a system of
interacting TLS's arranged in a linear  configuration and affected
by a bosonic environment, considering both the individual and
common-environment cases. By properly choosing the basis onto
which each TLS-oscillator interaction is described, we have been
able to push the analytic approach to such a problem up to the
point of constructing a handy recipe for tracking any regime of
interaction, from the weak to the strong-coupling regime with, in
principle, an arbitrary number of environmental modes per site.
This has paved the way to the study of the effects that quantum
entanglement shared by the elements of the environment has on the
dynamics of the TLS system. Entanglement appears to enhance the
non-Markovian character of the evolution, giving rise to specific
features in relevant figures of merit of a TLS chain that are not
present in the case of uncorrelated environmental states. We have
illustrated such effects by studying a dimer, which is the basic
building-block for the construction of interesting interaction
configurations that are usually considered for the description of
biological complexes. Our investigation leaves open a series of
questions related, in particular, to the scaling of the effects
revealed here with the degree of connectivity, the depth and the
geometry of the network of interacting TLS's. We plan to address
such issues by extending and improving the
implementability of our method.

\acknowledgments
MP thanks A. Rivas and S. F. Huelga for useful discussions. SL thanks the Centre for Theoretical Atomic, Molecular and Optical Physics, Queen's University Belfast, for hospitality during the completion of this work. We acknowledge financial support from the UK EPSRC (EP/G004759/1) and the British Council/MIUR British-Italian Partnership Programme 2009-2010.


\begin {thebibliography}{99}
\bibitem{LeggettReview} A. J. Leggett, S. Chakravarty, A. T. Dorsey, M. P. A. Fisher, A. Garg, and W. Zwerger, Rev. Mod. Phys. {\bf 59}, 1 (1987).
\bibitem{Reina}J. Eckel, J. H. Reina and M. Thorwart, New J. Phys {\bf 11}, 085001 (2009).
\bibitem{Huelga}F. Caruso, A. W. Chin, A. Datta, S. F. Huelga and M. Plenio, Phys. Rev. A {\bf 81}, 062338 (2010);
\bibitem{lloyd}P. Rebentrost, M. Mohseni, I. Kassal, S. Lloyd, and A. Aspuru-Guzik, New J. Phys. {\bf11 } 033003
(2009).
\bibitem{scholak} T. Scholak, {\it et al}., Phys. Rev. E {\bf 83},
021912 (2011).
\bibitem{Huelga2}F. Caruso, A. W. Chin, A. Datta, S. F. Huelga and M. Plenio, J. Chem. Phys. {\bf 131}, 105106
(2009).
\bibitem{nmqj} P. Rebentrost, R. Chakraborty, and A. Aspuru-Guzik, J. Chem. Phys. {\bf 131}, 184102 (2009).
\bibitem{diamond} P. E. Barclay, K.-M. Fu, C. Santori, and R. G. Beausoleil, Optics Express {\bf 17}, 9588 (2009);
Appl. Phys. Lett. {\bf 95}, 191115 (2009); P. E. Barclay, K.-M. C. Fu, C. Santori, A. Faraon, R. G. Beausoleil, arXiv:1105.5137; 
A. Faraon, P. E. Barclay, C. Santori, K.-M. C. Fu, and R. G. Beausoleil, Nat. Photon. {\bf 5}, 301 (2011); C. Santori, P. E. Barclay, K.-M. C. Fu, 
S. Spillane, M. Fisch, and R. G. Beausoleil, Nanotechnology {\bf 21}, 274008 (2010). 
\bibitem{cqed} A. Wallraff, {\it et al}., Nature (London) {\bf 431}, 162
(2004); M. Hofheinz, {\it et al}., Nature (London) {\bf 454}, 310
(2008); J. M. Fink, {\it et al}., Nature (London) {\bf 454}, 315
(2008); L. S. Bishop, {\it et al}., Nature Phys. {\bf 5}, 105
(2009); O. Astafiev, {\it et al.}, Science {\bf 327}, 840 (2010); J. M.
Fink, {\it et al}., Phys. Rev. Lett. {\bf 105}, 163601 (2010); C.
Lang, {\it et al}., Phys. Rev. Lett. {\bf 106}, 243601 (2011).
\bibitem{bourassa}
J. Bourassa {\it et al}., Phys. Rev. A {\bf 80}, 032109 (2009); A.
Fedorov, A. K. Feofanov, P. Macha, P. Forn-D\'{i}az, C. J. P. M.
Harmans, and J. E. Mooij, Phys. Rev. Lett. {\bf 105}, 060503
(2010); P. Forn-D\'{i}az, J. Lisenfeld, D. Marcos, J. J.
Garc\'{i}a-Ripoll, E. Solano, C. J. P. M. Harmans, and J. E. Mooij,
Phys. Rev. Lett. {\bf 105}, 237001 (2010); J. Casanova, G. Romero,
I. Lizuain, J. J. Garc\'{i}a-Ripoll, and E. Solano, Phys. Rev.
Lett. {\bf 105}, 263603 (2010).
\bibitem{massimo} G. M. Palma, K.-A. Suominen, and A. Ekert, Proc. R. Soc. Lond. A {\bf 452}, 567
(1996); J. H. Reina, L. Quiroga, and N. F. Johnson, Phys. Rev. A
{\bf 65}, 032326 (2002).
\bibitem{barnettradmore} S. M. Barnett and P. M. Radmore, {\it Methods in Theoretical Quantum Optics} (Oxford University Press, New York, 1997).
\bibitem{AS} M. Abramowitz and I. Stegun, {\it Handbook of Mathematical Functions with Formulas, Graphs, and Mathematical Tables} (Dover, New York, 1964).
\bibitem{comment} One can relate the Charlier polynomials to the associated Laguerre polynomials as $C_n(m; x^2)=(-1)^n n! L_n^{(-1-m)}\left(1/x\right)$, where $x$ is a generic argument.
\bibitem{logneg} M. B. Plenio, Phys. Rev. Lett. {\bf 95}, 090503 (2005).
\bibitem{meas1} M. M. Wolf, J. Eisert, T. S. Cubitt, and J. I. Cirac, Phys. Rev. Lett. {\bf 101}, 150402 (2008).
\bibitem{meas2} A. Rivas, S. F. Huelga, and M. B. Plenio, Phys. Rev. Lett. {\bf 105}, 050403
(2010); D. Chruscinski, A. Kossakowski, and A. Rivas, Phys. Rev. A
{\bf 83} 5 (2011).
\bibitem{meas3} H.-P. Breuer, E. M. Laine, and J. Piilo, Phys. Rev. Lett. {\bf 103}, 210401 (2009); E.-M. Laine, J. Piilo, and H.-P. Breuer, Phys. Rev. A {\bf 81}, 062115 (2010).
\bibitem{meas4} X.-M. Lu, X. Wang, and C. P. Sun, Phys. Rev. A {\bf 82}, 042103 (2010).
\bibitem{app1} T. J. G. Apollaro, C. Di Franco, F. Plastina, M. Paternostro, Phys. Rev. A {\bf 83}, 032103 (2011).
\bibitem{app2} P. Haikka, S. McEndoo, G. De Chiara, M. Palma, and S. Maniscalco, arXiv:1105.4790 (2011).
\bibitem{trineg} The tripartite negativity is defined as ${\cal N}_{tri}{=}[{\cal N}_{1|23}{\cal N}_{2|31}{\cal N}_{3|12}]^{1/3}$, with ${\cal N}_{j|jk}$ the negativity in the ${i-(j,k)}$ bipartition. 
\end{thebibliography}
\end{document}